\documentclass[prl,twocolumn,amsmath,amssymb,floatfix,letter]{revtex4-1}
\usepackage[dvips]{graphicx}
\usepackage{color,array,dcolumn}
\usepackage{amsmath}
\usepackage{amssymb}

\begin{document}

\title{Physical Adsorption at the Nanoscale: Towards Controllable Scaling of the Substrate-Adsorbate van der Waals Interaction}

\author{Alberto Ambrosetti$^{1}$}
\author{Pier Luigi Silvestrelli$^1$}
\author{Alexandre Tkatchenko$^{2}$}
\affiliation{
$^1$Dipartimento di Fisica e Astronomia, Universit\`{a} degli Studi di Padova, via Marzolo 8, \textsl{35131}, Padova, Italy \\
$^2$Physics and Materials Science Research Unit, University of Luxembourg, L-1511 Luxembourg}

\begin{abstract}
\date{\today}
The Lifshitz-Zaremba-Kohn (LZK) theory is commonly considered as the correct large-distance limit for the 
van der Waals (vdW) interaction of adsorbates (atoms, molecules, or nanoparticles) with solid substrates.
In the standard approximate form, implicitly based on {\it local} dielectric functions, the LZK approach 
predicts universal power laws for vdW interactions depending only on the dimensionality of
the interacting objects. However, recent experimental findings are challenging the universality of this 
theoretical approach  at finite distances of relevance for nanoscale assembly. 
Here, we present a combined analytical and numerical many-body study demonstrating that physical adsorption 
can be significantly enhanced at the nanoscale. Regardless of the band gap or the nature of the adsorbate specie,
we find deviations from conventional LZK power laws that extend to separation distances of up to 10--20 nanometers.
Comparison with recent experimental observation of ultra long-ranged vdW interactions in the delamination of graphene 
from a silicon substrate reveals qualitative agreement with the present theory. 
The sensitivity of vdW interactions to the substrate response and to the adsorbate characteristic excitation 
frequency also suggests that adsorption strength can be effectively tuned in experiments, paving the way to an 
improved control of physical adsorption at the nanoscale.
\end{abstract}

\maketitle


Non-covalent van der Waals (vdW) interactions constitute a universal cohesive force whose impact
extends from the atomistic scale~\cite{Stone-book,Kaplan-book} to a wealth of macroscopic phenomena 
observed on a daily basis~\cite{Langbein-book,Parsegian-book}.
With an influence ranging from protein-drug binding to the double helix in DNA~\cite{PNAS-2012},
the peculiar pedal adhesion in the gecko~\cite{AutumnN00,AutumnPNAS02},
and even cohesion in regolith and rubble-pile asteroids~\cite{asteroid1,asteroid2},
these non-bonded forces are quantum mechanical in origin and arise from electrodynamic interactions
between the constantly fluctuating electron clouds that characterize molecules
and materials~\cite{science}. While our understanding of vdW interactions is rather complete
at the smallest atomistic and the largest macroscopic scales, these pervasive forces exhibit 
a range of surprising and poorly understood effects at the 
nanoscale~\cite{Dobson,Dobson-manybody,chang,Misquitta,Misquitta2,Kotov-science,science}. 

This lack of understanding is best exemplified by recent puzzling experimental observations, which 
include: (\textit{i}) ultra long-range vdW interactions extending up to tens of nm into heterogeneous 
Si/SiO$_2$ dielectric interfaces~\cite{Jacobs1,Jacobs2}, and influencing the delamination of extended 
graphene layers from silicon substrate~\cite{grafsurf},
  (\textit{ii}) complete screening of the vdW 
interaction between an atomic force microscope (AFM) tip and a SiO$_2$ surface by the presence of a 
single layer of graphene adsorbed on the surface~\cite{Tsoi-ACSNano}, (\textit{iii}) super-linear 
sticking power laws for the physical adsorption of metallic clusters on carbon nanotubes with 
increasing surface area~\cite{Khlobystov-ACSNano}, and (\textit{iv}) non-linear increases in 
the vdW attraction between homologous molecules and an Au(111) surface as a function of the 
molecular size~\cite{Tautz-NatureComm}.
Recently, theoretical evidence was found for exceptionally long-ranged vdW attraction
between coupled low-dimensional nanomaterials with metallic character~\cite{Dobson} or
small band-gap~\cite{Misquitta,science}. Observed major deviations from conventional 
pairwise predictions~\cite{science} stem from non-local dipolar fluctuations
induced by the low dimensionality of the structures~\cite{Louie}. 

While analysis of these striking phenomena focused on coupled 1D and 2D nanomaterials, 
the broader and technologically relevant problem of physical adsorption of atoms, molecules 
or nanoparticles on low-dimensional structures is not yet fully understood.
This lack of comprehension is mostly related to the intrinsically local charge fluctuations 
of small adsorbates and the non-negligible 
HOMO-LUMO gaps, which may suggest {\it weak coupling} to the {\it soft} delocalized polarization 
modes of the substrate. However, transient electronic excitations in low-dimensional substrates could 
cause unexpectedly strong electrodynamic fields, whose effects are yet to be assessed.

Both energetics and dynamical properties of physically adsorbed moieties can largely depend on the precise vdW scaling. 
Experimental implications of possible unexpected trends in nanoscale physical adsorption 
can thus range from catalysis and wetting to film deposition or self-assembly.
State-of-the-art single-molecule AFM experiments are now also able to measure power law exponents 
governing the adsorption energy of large molecules on solid substrates
to a precision of $\pm$0.2~\cite{Tautz-NatureComm}. 
Such experimental progress provides a substantial challenge for the theoretical understanding of vdW interactions
and precise modeling of their effects at the nanoscale. 
To achieve both goals, here we utilize a combined analytic and numerical many-body model of physical adsorption to
systematically study the interaction of adsorbates with a range of both metallic and finite-gap 
low-dimensional 1D and 2D substrates. 
Even for the smallest atomic adsorbates, we find that the strongly non-local 
response of these substrates, which stems from
coherent wave-like electronic fluctuations, causes qualitative deviations from
conventional vdW energy predictions. In fact, the vdW adsorption energy can exhibit a peculiar 
slow decay over length scales extending from $\sim$ 5 \AA\, to well above 10 nm.
Interestingly, the interaction energy decay can be further regulated by a suitable choice of the substrate 
response and of the adsorbate moiety, thus opening a plethora of pathways towards detailed and 
selective experimental control of vdW forces at the nanoscale.

So far, the theoretical modeling of the complex many-body vdW interactions arising on extended substrates
has mostly relied on the Lifshitz-Zaremba-Kohn (LZK) theory~\cite{Lifshitz,Zaremba,mostepanenko-rev}.
In principle the LZK approach provides an exact theoretical framework, where explicit dependence on the {\it interacting} substrate 
response function $\chi_{\rm S}$ ensures complete inclusion of many-body screening effects. 
Due to the intrinsic complexity of $\chi_{\rm S}$, however, the {\it interacting} susceptibility is normally approximated in LZK
calculations by an implicitly {\it local} form.
Essentially, by approximating the substrate response in terms of the {\it average} dielectric function 
(computed at wavevector $\mathbf{q}$=0) the complexity of the problem can be strongly reduced, and simple power law 
expressions can be derived for the vdW interaction energy $\Delta E_{\rm vdW}$. 
This is exemplified for instance by the well known expression
$\Delta E_{\rm vdW}\sim C_3/D^3$, derived for small molecular fragments at large distance $D$ from a 
semi-infinite substrate, and extended (with different power law dependence on $D$), to treat also
lower-dimensional substrates~\cite{Cole2,bordag}.
Within the LZK approach the overall Hamaker constants ($C_3$ in the above expression)  
are renormalized with respect to standard pairwise vdW approximations~\cite{Grimme-D3,TS-vdW,LangLunq,WannierC6,BeckesC6}, 
due to the effective inclusion of screening effects in the extended substrate.
However, the vdW interaction power laws predicted in the {\it local} LZK limit exactly 
coincide with those of additive pairwise vdW approaches. 
While this approach is generally correct for bulk-like substrates, here we will analyze in detail the implications of the 
{\it locality approximation}, evidencing major shortcomings in the rapidly emerging context of low dimensional substrates. 
By explicitly accounting for non-local electron charge fluctuation we will thus provide a correct application of the LZK theory to 
substrates with arbitrary dimensionality and response properties.

\begin{figure}
\includegraphics[width=8.0cm]{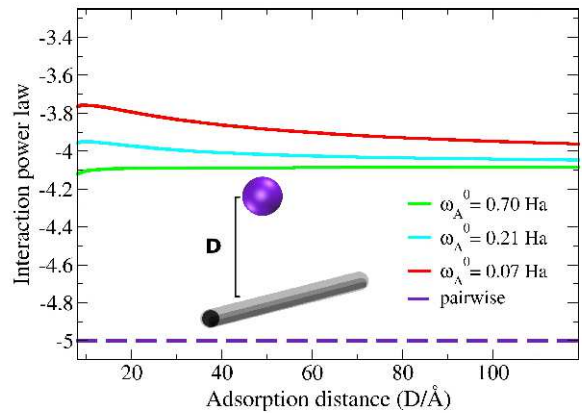}
\caption{Power law decay of the vdW interaction between a single adsorbate and an infinite metallic wire. Adsorbates with different characteristic frequency $\omega_{\rm A}^0$ are considered in order to visualize the interdependence between the adsorbate dynamical polarizability and the interaction power law. Comparison with the standard pairwise power law ($D^{-5}$) indicates more evident deviations from the pairwise behavior in adsorbates with lower characteristic frequency. We also note from Eq.~\eqref{evdw1} that the static polarizability $\alpha_{\rm A}^0$ can be factorized, hence only providing a constant overall rescaling of the vdW interaction.}
\label{fig1}
\end{figure}


In order to introduce the essential physical concepts, we begin our analysis by considering a single
adsorbate $A$ (for instance an atom, a molecule, or a nanoparticle) interacting with a one-dimensional (1D) 
metallic wire $W$ at a separation $D$. 
Atomic units ($e=m=4\pi\epsilon_0=\hbar=1$) are adopted hereafter to simplify the notation. 
The wire density-density response $\chi^{\rm RPA}_{\rm W}$  
can be computed starting from the 1D free electron gas bare susceptibility~\cite{longe,vignale} 
$\chi^0_{\rm W}(q,\omega)=N_0q^2/\omega^2$, where $N_0$ is the number of electrons per unit length, 
and the intra-wire Coulomb interaction~\cite{Dobson} $v_{\rm W}(q)=-2e^2ln(qb)$
 ($b$ being the effective wire thickness, in the limit $bq<<1$).
The vdW interaction energy is thus evaluated by coupling
the polarizability $\alpha_A(i\omega)$ of the adsorbate 
to the RPA interacting substrate response through the wire-adsorbate interaction~\cite{Dobson} 
$2K_0(qD)$ (a modified Bessel function of the second kind, see Ref.~\cite{Zaremba}), as:
\begin{equation}
\label{evdw0}
\Delta E_{\rm vdW}=\int_0^{\infty} d \omega \int dq\, \frac{q^2 I(qD)}{\pi^2} \chi^{\rm RPA}_{\rm W}(q,i\omega) \alpha_{\rm A}(i\omega)\,.
\end{equation}
Here $I(qD)=\left( K_0(qD)^2+ K_0'(qD)^2 \right)$, and  momentum integration is restricted to the first Brillouin zone.
For $D$ larger than the adsorbate characteristic dimension, 
the dipole approximation can be adopted for the response of the adsorbate $A$.
We also map now the adsorbate polarizability onto the response of a single quantum harmonic oscillator making use of the single Lorentzian expression
$\alpha_{\rm A}(i\omega)=\alpha^0_{\rm A}/(1+(\omega^2/\omega^0_{\rm A})^2)$ (being $\alpha^0_{\rm A}$ the static polarizability 
and $\omega^0_{\rm A}$ the characteristic oscillator frequency).
This procedure corresponds to introducing a single effective excitation mode for the adsorbate, and has been widely 
applied in literature~\cite{Cole2,botti2011,Zaremba-book}.
Alternative treatments based on multiple excitation modes, however, are equally possible within this 
framework, and can be reformulated in terms of linear combinations of the single mode contributions considered hereafter.
After analytical frequency integration, the following expression is obtained:
\begin{equation}
\Delta E_{\rm vdW}=-\int dq\, \frac{q^2 I(qD)}{2\pi} \frac{\alpha_A(0)\,\omega^0_{\rm A}\, q}{L(q) (\omega^0_{\rm A}+qL(q))}\,,
\label{evdw1}
\end{equation}
where $L(q)=\sqrt{2|\ln(qb)|}$.
We note that, at variance with conventional LZK theory, the explicit $q$ dependence of the substrate response function is preserved in this derivation, 
thus accounting for the actual non-locality of the charge fluctuation modes.

To study the scaling of $\Delta E_{\rm vdW}$ with respect to the adsorption distance, 
we first observe that the rapidly decaying interaction factor $I(qD)$ introduces an effective
integration cutoff at $q\sim 1/D$. By performing the variable substitution $q'=qD$, it becomes thus evident that the power law 
scaling of $E_{\rm vdW}$ is determined by the $q$ dependence of the integrand~\cite{Dobson,science}, 
and it specifically varies depending on the relative magnitude of the terms at the denominator 
(namely $\omega^0_{\rm A}$ and $qL(q)$). 
In particular, we can distinguish two separate regimes:
{\bf i)} for $\omega^0_{\rm A}>>L(1/D)/D$ 
the integrand is roughly proportional to $I(qD)q^{3}/L(q)$ over the whole integration domain. 
Integration over $q$ thus leads to $E_{\rm vdW}\sim D^{-4}$ up to logarithmic corrections.
{\bf ii)} if the opposite case holds ({\it i.e.} $\omega^0_{\rm A}<<L(q)q$ over most of the integration domain) then the integrand
 becomes roughly proportional to $I(qD)q^{2}/L^2(q)$, leading to power law scalings that are 
intermediate between $\sim D^{-3}$ and $\sim D^{-4}$.
According to the above analysis, by increasing $D$ regime {\bf i)} is eventually approached, 
and the transition between regimes {\bf ii)} and {\bf i)} is influenced by the
adsorbate characteristic frequency $\omega^0_{\rm A}$: in fact, for small values of $\omega^0_{\rm A}$
the $\sim D^{-4}$ scaling is approached at larger adsorption distances  (see Fig.~\ref{fig1}). 
At high $\omega_{\rm A}$, instead, regime {\bf i)} is soon approached, but the power law can show a slight initial growth due to the logarithmic corrections, remaining however close to the $\sim D^{-4}$ asymptote.

Based on the above analysis, we note that the vdW interaction energy between substrate and adsorbate
exhibits evident qualitative deviations from conventional pairwise predictions ($D^{-5}$ in 1D), 
implying an ultra-slow decay of the interaction with respect to the adsorption 
distance. The existence of separate scaling ranges, moreover, suggests that the 
interaction details may be experimentally tuned by an appropriate choice of the adsorbate species:
physically speaking, adsorbates with different $\omega^0_{\rm A}$ will be sensitive to 
different frequency ranges, and, correspondingly, to distinct characteristic modes of the substrate.
Similar conclusions can be drawn by considering the tight-binding response function for 1D metallic chains
proposed by Misquitta {\it et al.}~\cite{Misquitta}. In that case, however,
the non-locality of $\chi_W$ entirely derives from the collective character of the quasi-particle 
eigenstates, and does not stem from the self-consistent RPA treatment of the Coulomb coupling. 

While the above results are specifically derived for a metallic 1D {\it substrate}, we extend now our 
treatment to finite-gap structures. Due to electronic charge localization one can describe in this case
the response of a $N$-atom system  in terms of $N$ interacting atomic polarizabilities. These can be mapped
onto a set of coupled atom-centered quantum harmonic oscillators, as outlined by the MBD framework~\cite{MBD,science},
by introducing the coupled dipolar Hamiltonian~\cite{Bade,Donchev}:
\begin{equation}
\label{mbdhamiltonian}
H_{\rm MBD}=-\sum_{p=1}^N \frac{\nabla^2_{\boldsymbol{\mu}_p}}{2}+ \sum_{p=1}^N \frac{\omega_p^2 \boldsymbol{\mu}_p^2}{2} + \sum_{p\neq q }^N \omega_p \omega_q \sqrt{\alpha^0_p\alpha^0_q}\boldsymbol{\mu}_p T_{pq} \boldsymbol{\mu}_q \,.
\end{equation}
The $p$-th atom is characterized by the static polarizability $\alpha^0_p$ and the characteristic 
frequency $\omega_p$, and $\boldsymbol{\mu}_p$ describes the mass-weighted charge displacement from the
ionic position $\mathbf{R}_p$~\cite{Donchev}. 
The interaction tensor $T$ introduces a dipolar coupling between different oscillators, and is defined as
$T_{pq}=\nabla_{\mathbf{R}_p} \nabla_{\mathbf{R}_q}v(R_{pq})$, where $v(R_{pq})$ is the Coulomb 
interaction between atoms $p$ and $q$, damped at short range due to gaussian charge overlap~\cite{ACFDT-2013}.
Given the quadratic dependence on $\boldsymbol{\mu}_p$, $H_{\rm MBD}$ can be exactly diagonalized, 
leading to a set of $3N$ {\it interacting frequencies} $\bar{\omega}_p$, from which the vdW 
energy can be promptly computed as $E_{\rm vdW,MBD}=\left(3\sum_{p=1}^{N}\omega_p-\sum_{p=1}^{3N}\bar{\omega}_p\right)/2$. 
The resulting dispersion energy is mathematically equivalent to the RPA~\cite{ACFDT-2013} long-range
correlation energy arising between dipolar oscillators.

\begin{figure}
\includegraphics[width=8.0cm]{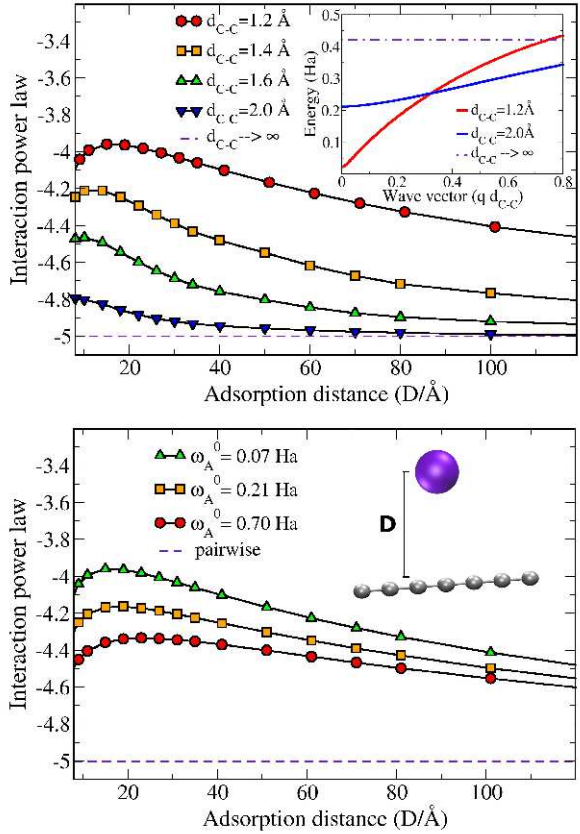}
\caption{MBD power law decay of the vdW interaction between single adsorbate (with characteristic frequency $\omega_{\rm A}^0=0.07$ Ha) and a periodic 1D atomic chain. Upper panel: chains with different interatomic distance $d_{\rm C-C}$ are considered. The larger deviations from the pairwise limit $D^{-4}$ observed at small $d_{\rm C-C}$ find a correspondence in the steep mode dispersion and small energy gap reported in the inset. In fact, the steep dispersion is indicative of a strongly non-local dipolar response in the chain.  Lower panel: dependence of the interaction power law on the adsorbate frequency ($\omega_{\rm A}^0$). In analogy to the metallic case, adsorbates with lower $\omega_{\rm A}^0$ exhibit more evident deviations from the pairwise $D^{-5}$ limit.}
\label{fig2}
\end{figure}

In order to investigate adsorption on 1D non-conducting systems, we consider a carbyne-like wire,
consisting of a linear chain of C atoms with equal nearest-neighbor distances $d_{\rm C-C}$. 
Different values of $d_{\rm C-C}$ are then analyzed, in order to assess the role of the chain response.
In Fig.\ref{fig2}, we observe that, also in finite-gap systems, sizeable deviations from the
the conventional $D^{-5}$ scaling are possible even beyond $\sim$10 nm.
In analogy with the case of two parallel 1D chains~\cite{science}, the power law initially increases reaching
a plateau, and subsequently decreases, gradually tending to the pairwise limit. Again, the effect
is enhanced by low adsorbate characteristic frequencies. Moreover,
power law deviations become evidently more pronounced at smaller $d_{\rm C-C}$ values,
while the vdW energy scaling rapidly approaches $D^{-5}$ beyond $d_{\rm C-C}=2.0$\AA.
The ultra-slow decay of vdW interactions is thus closely related to the non-locality of
the dipolar response of the chain. 
In fact, as observed in Ref.~\cite{science}, highly collective dipole-fluctuation modes can emerge in
low-dimensional structures. Such modes correspond to the dipole waves sustained by the 
system, and characterize the degree of non-locality of the response function. 
Within the MBD approach these collective wave-like modes are directly obtained as the eigenmodes of the Hamiltonian
\eqref{mbdhamiltonian}, and, depending on the dispersion of the corresponding eigenenergies ($\bar{\omega}(q)$),
different power law scalings of the vdW energy can be found.

This concept can be formalized by expressing the vdW adsorption energy $\Delta E_{\rm vdW,MBD}$ in integral form.
We thus take the continuum limit (valid at large $D$), and consider only the {\it longitudinal}
dipole fluctuation modes occurring in the chain. Transverse modes, in fact, provide smaller contributions 
to $\Delta E_{\rm vdW,MBD}$ at large $D$ and will be neglected for simplicity. 
Making explicit use of the {\it f}-sum rule~\cite{Pines-book}, we can express the polarizability of the collective
mode corresponding to wave vector $q$ as $\bar{\alpha}(q)(i\omega)=\bar{\alpha}^0(q)/(1+\omega^2/\bar{\omega}^2(q))$, where
the static polarizability is written in terms of the C static polarizability $\alpha_{\rm C}^0$ and the
characteristic frequency $\omega^0_{\rm C}$ as $\bar{\alpha}^0(q)=\alpha^0_{\rm C} (\omega^0_{\rm C}/\bar{\omega}(q))^2$.
By extending  Eq.\eqref{evdw1} to the present model, we can thus express the interaction energy as
\begin{equation}
\Delta E_{\rm vdW,MBD}=
-\int dq\, \frac{I(qD)}{2\pi} \frac{\alpha_{\rm A}(0)\alpha^0_{\rm C}\omega^0_{\rm A}(\omega^0_{\rm C}q^2)^2}{\bar{\omega}(q) (\bar{\omega}(q)+\omega^0_{\rm A})}\,.
\label{evdwmbd}
\end{equation}
In analogy with Eq.\eqref{evdw1}, the mode dispersion $\bar{\omega}(q)$ entering at the denominator ultimately determines the
power law scaling of $\Delta E_{\rm vdW,MBD}$. 
For instance, at large $d_{\rm C-C}$ the atoms in the chain
become weakly interacting, leading to flat energy dispersion ($\bar{\omega}(q)\sim const.$).
In presence of energetic degeneracy, localized dipole fluctuations can thus occur in the system
and the pairwise approximation becomes valid.
At realistic interatomic distances, instead, the intra-chain interaction acts by
lifting the modes degeneracy, leading to non-trivial mode dispersion (see Fig.\ref{fig2}). 
In particular, one observes that $\bar{\omega}(q)$ can assume very small values for $q\rightarrow0$~\cite{science}, showing then
a steep increase at growing $q$. The steep dispersion of the charge fluctuation modes is a clear signature of response non-locality (see Supplementary Material). 
We stress, though, that due to the intrinsic localization of the single QHOs (justified by the electron charge 
localization), $\bar{\omega}(0)$ is always non-zero. This property determines a qualitative asymptotic difference
with respect to the metallic case, implying that the $D^{-5}$ power law is recovered as the asymptotic limit.


To extend our treatment beyond 1D substrates, we now consider the adsorption on a two-dimensional
(2D) graphene substrate. Given the complexity of the full electronic structure, we make use of 
approximate response functions, based on the low energy excitations of the $\pi$ electrons.
Although this approximation neglects polarization components orthogonal to the plane, nonetheless it permits to
unravel the effects of the band structure near the Dirac cone, that govern the non-trivial electronic properties of graphene.
Besides the conventional RPA~\cite{Dobson,guinea} response function, a more accurate approximation is also
considered, derived including vertex corrections through a renormalization group (RG) approach~\cite{sodemann,Dobson-manybody}.
Details on the response functions, and on the computation of $\Delta E_{\rm vdW}$ using this approach are reported in the Supplementary Material.
From Fig.~\ref{fig3}, a clear analogy emerges between adsorption on 1D systems and graphene. 
Sizeable power law deviations from the pairwise limit extend beyond 100 \AA, and are again enhanced in presence
of adsorbates with low characteristic frequency.
Moreover, the semi-quantitative agreement existing between power laws derived within RPA and RG suggests that ring diagrams
can already account for relevant response delocalization, providing hence further support to the present MBD results.

In order to unravel how vdW interactions depend on the substrate details, 
we apply the MBD method to single-layer MoS$_2$ 
and finite-gap graphenic materials, setting the atomic polarizability $\alpha^0_{\rm C}$ to different values. 
Interestingly, by inspection of Fig.~\ref{fig3}, we find qualitative agreement between atomistic MBD calculations 
and the previous semi-analytical model. 
In addition, we observe that the more polarizable substrates~\cite{notemos2} are characterized by a slower decay of the adsorption
energy with respect to $D$. The analogies existing between higher $\alpha^0_{\rm C}$ and smaller $d_{\rm C-C}$ (see Fig.~\ref{fig2}),
can be understood considering that many body effects in MBD are effectively controlled by the dimensionless
quantity $\alpha^0_{\rm C}/d_{\rm C-C}^3$. An inverse proportionality thus exists between power law variations
induced by changes in the two quantities. The important deviations found for MoS$_2$ also suggest that other quasi-metallic or 
finite-gap low dimensional materials, such as transition metal dichalcogenides, silicene or phosphorene should
exhibit analogous trends in physical adsorption processes.

Going from monolayer graphenic structures to multilayer graphene (see Supplementary Material),
pairwise power laws are gradually approached at short $D$, suggesting that the conventional vdW asymptotic
decay should be typically recovered in bulk systems. At the same time, however, deviations from pairwise power laws 
become longer-lasting with respect to $D$ when increasing the number of layers: at large $D$ thin multilayered
structures effectively behave as a single layer with enhanced polarizability-to-surface ratio, thus inducing 
longer-ranged vdW interactions which require an appropriate description beyond the local LZK limit.

We finally note that extremely long-ranged interlayer forces have been observed in a very recent
 experiment~\cite{grafsurf} conducted by separating graphene from the native oxide layer on a Si(111) substrate by lateral wedge insertion and crack opening.
While vdW interactions are expected to contribute up to the $\sim$10 nm scale, the experimentally estimated 
delamination resistance per unit area only converged to a constant value at $\sim$1 $\mu$m crack openings. 
To interpret this puzzling result we considered cracks with longitudinal extent $a$ and quadratic increase 
of the graphene-Si(111) 
separation $h(x)$ with respect to the crack coordinate $x\in[0,a]$ (see Supplemetary material). 
By assuming a $\sim h(x)^{-2.5}$ power law scaling of the interaction (a variation of 0.5 from the pairwise power law is compatible 
with our findings), we found that the dispersion energy cost for crack 
formation (per unit area) approaches its converged value within 2\% only beyond  $h(a)\sim$0.5 $\mu$m.
Considering instead cracks with constant opening $h$ and a $h^{-3}$ interaction scaling, the dispersion
energy cost is converged within 2\% already at $h\sim$2 nm.  Our simple analysis can thus qualitatively capture the 
observed ultra long-ranged sticking effect. 
Moreover, the combined many-body polarization enhancement in graphene~\cite{Vivek}, and the complex strain effects occurring
in the system upon mechanical deformation could further extend the effective range of the effective interlayer interaction.


In conclusion, we evidenced highly non-trivial power law scalings of the vdW interaction arising between 
atoms or small molecules and both metallic and finite-gap low-dimensional substrates. These power laws substantially deviate from 
standard pairwise predictions, and result in ultra long-ranged dispersion forces.
This effect arises due to marked non-localities of the substrate response, and could only be captured by accounting 
for the detailed momentum dependence of the susceptibility within a full many-body approach.
The non-trivial dispersion enhancements predicted for atomic adsorbates demonstrate that any type of system --from the atomistic 
scale up to the nanoscale-- can undergo ultra-long ranged vdW forces in the presence of polarizable low dimensional substrates.
The sensitivity of the vdW energy scaling to the adsorbate characteristic frequency and substrate response properties paves 
the way to a detailed and selective control of molecule-substrate interactions. 
These results open new perspectives for challenging experimental manipulations of adsorption and nanoassembly phenomena.
Possible implications may also extend to the broad context of low-dimensional biological systems, including phospholipid aggregates and 
bilayers~\cite{phospholipids}, or extended polypeptide chains~\cite{blum}.

A. Ambrosetti acknowledges useful and insightful discussion with F. Toigo.

\begin{figure}
\includegraphics[width=8.0cm]{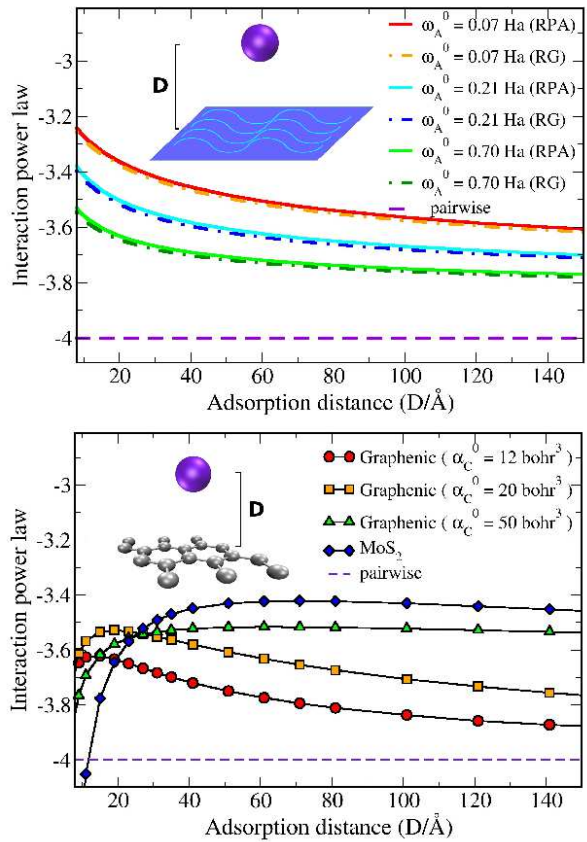}
\caption{Power law decay of the vdW interaction between single atoms and 2D materials. Upper panel: $\Delta E_{vdW}$ computed at different adsorbate characteristic frequencies, adopting analytical RPA and RG response functions for graphene. Lower panel: MBD results for 2D materials characterized by graphenic structure and variable atomic polarizability ($\omega^0_{\rm A}=0.07$ Ha). The pairwise asymptotic limit $D^{-4}$ is reported for comparison.}
\label{fig3}
\end{figure}

\bibliography{literature}

\begin{thebibliography}{48}
\bibitem [{not()}]{notemos2}%
  \BibitemOpen
  \href@noop {} {}\bibinfo {note} {{According to the Tkatchenko-Scheffler
  parameterization, MoS$_2$ is characterized by an average polarizability per
  unit area which is about 4 times larger than in graphene.}}\BibitemShut
  {Stop}%
\end{thebibliography}

\end{document}